# Magnetodynamic properties of dipole-coupled 1D magnonic crystals

*Suraj Singh[1*], Xiansi Wang[1], Ankit Kumar[2], Alireza Qaiumzadeh[1], Peter Svedlindh[2], Thomas Tybell,[3] and Erik Wahlström[1]*

[1]*Center for Quantum Spintronics, Department of Physics, NTNU - Norwegian University of Science and Technology, NO-7491 Trondheim, Norway*

[2]*Department of Materials Sciences and Engineering, Uppsala University, Box 516, SE-75121 Uppsala, Sweden*

[3]*Department of Electronic Systems, NTNU - Norwegian University of Science and Technology, NO-7491 Trondheim, Norway*

[*]*Corresponding author e-mail: rsinghsuraj1992@gmail.com*

**Abstract**

Magnonic crystals are magnetic metamaterials, that provide a promising way to manipulate magnetodynamic properties by controlling the geometry of the patterned structures. Here, we study the magnetodynamic properties of 1D magnonic crystals consisting of parallel NiFe strips with different strip widths and separations. The strips couple via dipole-dipole interactions. As an alternative to experiments and/or micromagnetic simulations, we investigate the accuracy of a simple macrospin model. For the case of simple strips, a model with a single free parameter to account for an overestimation of the out of plane demagnetization of the magnonic lattice is described. By adjusting this parameter, a good fit with experimental as well as micromagnetic results is obtained. Moreover, the Gilbert damping is found independent of lattice constant however the inhomogeneous linewidth broadening found to increase with decreasing stripe separation.

**Keywords:**     Permalloy, Magnonic Crystals, Dipole coupling, Ferromagnetic resonance

## 1. Introduction

Spin dynamics in nanostructured materials have attracted attention due to interesting underlying physics and potential for technological applications. Magnonic crystals (MCs) are a class of artificial magnetic media that offer a promising way to manipulate the magnetodynamic properties in microwave frequency by exploiting the patterned geometry[1,2]. Due to their interesting magnetic properties, MCs find applications in a wide range of magnetic devices such as advanced magnetic storage, data processing, and spin logic gates[3,4,5]. As a consequence, MCs have been studied extensively both theoretically and experimentally in numerous magnetic systems in order to explore the impact of MCs control parameters on its static and magnetodynamic properties, and for their potential application in novel magnonic devices[6-8,9,10,11,12,13,14].

Advances in lithography techniques make it possible to fabricate nanometer-sized MCs with narrow spacings. 1D MCs, with a periodic magnetic strip pattern along one direction have attracted considerable attention due to their simple geometry, convenient for studying the impact of lattice confinement on the magnetodynamic properties at the nanoscale [7,10,6]. In such systems, the dipolar coupling of magnetic strips plays an important role in the magnetodynamic properties. When the strips form a closely packed array, the fundamental mode of individual strips couples via a dynamic dipolar interaction resulting in formation of collective spin-wave excitations[6,7,15,16,17,18]. This is the result of the dynamic dipolar magnetic field removing the degeneracy between the discrete energy levels of the different magnetic elements. The collective dynamics stemming from the magnetodynamic dipolar interaction affect the writing time in closely packed storage media, the synchronization of spin-torque oscillators and most importantly the spin-wave dynamics in MCs[19]. The spacing between adjacent magnetic elements in such systems is a central parameter governing the interstrip dipolar coupling. Thus, investigating the effect of dipolar coupling on the magnetodynamic properties gives valuable information on the underlying physics and for potential application of MCs in magnonic devices.

Most of the previous investigations have been focused on Brillion Light Scattering (BLS) studies of dipolarly coupled 1D MCs, where the interplay of dipolar coupling on collective mode excitations and the formation of magnonic bandgaps have been studied extensively[7,11,16]. Ferromagnetic resonance (FMR) is a sensitive nondestructive technique allowing to study the magnetodynamic properties of the MCs. However, there are only a few reports of FMR studies of dipole-coupled MCs [20,21], which is especially true for detailed studies of important control parameters such as size and separation of the building blocks making up the MCs and their impact on the spin dynamics. Also, the impact of dipolar coupling on magnetic damping, which is important to lower the power of magnonic devices, is poorly understood. In this paper, we present a study of magnetodynamic properties of dipolarly coupled 1D MCs by FMR spectroscopy. The 1D MCs consist of parallel Permalloy (Py) strips prepared using electron beam lithography. We report the effect of strip width and lattice constant on the resonance field and describe a simple macrospin model that can be used to predict resonance behavior of 1D MCs. Also, the impact of the MC structure on the FMR linewidth has been investigated by broadband FMR spectroscopy.

## 2. Experimental Details

The MCs consisted of Py ($Ni_{80}Fe_{20}$) deposited on silicon substrates by e-beam evaporation. Electron beam lithography and lift-off techniques were used to fabricate the 1D strip-based MCs, having variable strip width $w$ and inter-strip separation $s$. The lattice constant of the MCs is $\lambda = w + s$. Fig. 1 (a) shows the SEM image of a sample described by $\lambda = 100$ nm and $w = 50$ nm. The total area of each MC is 1×1 mm². A constant deposition rate was used for all samples to ensure approximately the same thickness, $d$ of $14 \pm 3$ nm.

The magnetodynamic properties of the MCs were investigated by two complementary FMR techniques. Cavity FMR measurements were carried out in a commercial *X*-band electron paramagnetic resonance (EPR) setup with a fixed microwave frequency of 9.4 GHz (Bruker Bio-spin ELEXSYS 500, with a cylindrical TE-011 microwave cavity). The setup is equipped with a goniometer allowing to rotate the sample 360° in-plane as well as out-of-plane. A schematic of the sample rotation including the magnetization and magnetic field vectors is shown in fig. 1 (b). A microwave field is applied to the cavity and an applied dc magnetic field is swept to record the microwave absorption. The measurements were performed with low amplitude modulation of static field with lock-in detection to enhance the signal to noise ratio. To extract the resonant field, the measured FMR absorption was fitted to a sum of the derivative of symmetric and antisymmetric Lorentzian functions, and the line-shape parameters such as resonant field and linewidth were extracted[22].

For broadband FMR measurements, a microwave signal generator FMR setup with a coplanar waveguide (CPW) and lock-in amplifier detection technique was employed. A pair of homemade Helmholtz coils generating a low-frequency (211.5 Hz) and low-amplitude magnetic field (0.25 mT) was used to modulate the microwave signal, which was detected by the lock-in amplifier. The FMR spectra were recorded sweeping the dc magnetic field at constant microwave frequency. The measurements were taken at various frequencies ranging from 5 to 16 GHz in steps of 0.5 GHz with the dc magnetic field applied parallel and perpendicular to the magnetic strips of the MCs[23].

## 3. Results and Discussion

### 3.1 Resonant field

To investigate the magnetodynamic properties, we studied MCs with lattice constants ranging from 100 nm to 550 nm with a fixed width $w = 50$ nm. In this subsection, all the experiments were done using a 9.4 GHz cavity. The obtained resonant fields as a function of the in-plane rotation angle for different lattice constants are shown in fig. 2(a). For each sample, two modes can be observed for a magnetic field applied perpendicular to the strips - along the $x$-axis ($\phi = 0°$). The two modes shift towards each other as the applied field direction rotates away from $\phi = 0°$, and at $\phi = \pm 15°$ the modes merge into a single-mode before disappearing. The two modes correspond to two equilibrium magnetization directions. No modes are observed for a wide range of field directions around the $y$-direction i.e., along the strips. The frequency of the easy-axis mode falls outside the detectable range of cavity FMR measurements. The resonant field decreases with decreasing $\lambda$, which is due to the increasing dipolar interactions. We then fix the magnetic field along $\mathbf{H} \parallel \hat{x}$ ($\phi = 0°$) and plot the higher resonant field versus the lattice constant $\lambda$ in fig. 2(b). The resonant field increases with increasing $\lambda$.

To understand the observed magnetodynamic behaviors, we develop a macrospin analytical model for MCs, and verify its validity by micromagnetic simulations. Each magnetic strip of the MCs is considered as a macrospin with synchronized precession of the spins. The magnetodynamics of the MC is governed by the Landau-Lifshitz-Gilbert (LLG) equation,

$$\frac{\partial \mathbf{m}}{\partial t} = -\gamma \mathbf{m} \times \mathbf{H}_{\text{eff}} + \alpha \mathbf{m} \times \frac{\partial \mathbf{m}}{\partial t},$$

where $\mathbf{m}$ is the unit vector along the magnetization direction of each strip, $\gamma = \frac{g\mu_B}{\hbar}$ is the gyromagnetic ratio, $g$ is the Landé g-factor, $\mu_B$ is the Bohr magneton, and $\hbar$ is the reduced Planck's constant. There are different values of $\gamma$ corresponding to $g$ =2.00 to 2.17 used in literature[24]. Here we use $\gamma$ =176 Grad/s/T corresponding to $g$ =2.00. The specific value of $\gamma$ does not qualitatively affect the physics we discuss. The term $\alpha$ is the Gilbert damping parameter and $\mathbf{H}_{\text{eff}}$ is the total effective field,

$$\mathbf{H}_{\text{eff}} = H\hat{\mathbf{x}} - M_s(N_x m_x \hat{\mathbf{x}} + N_y m_y \hat{\mathbf{y}} + N_z m_z \hat{\mathbf{z}}) + \mathbf{H}_{\text{int}},$$

where $H\hat{\mathbf{x}}$ is the applied external field along the $x$-direction and $-M_s(N_x m_x \hat{\mathbf{x}} + N_y m_y \hat{\mathbf{y}} + N_z m_z \hat{\mathbf{z}})$ is the demagnetization field of the strip with $M_S$ being the saturation magnetization and $N_{x,y,z}$ the demagnetization components of each strip which can be calculated analytically from the strip dimensions (width $w$, thickness $d$ and the length of the strip[25]). The magnetic damping was neglected when calculating the resonant field of the modes for the sake of simplicity. $\mathbf{H}_{\text{int}}$ is the inter-strip dipolar interaction field. To calculate $\mathbf{H}_{\text{int}}$, we consider a strip with two neighbouring strips as illustrated in fig. 2(c) and calculate the dipolar field from the neighbouring strips at the center point of the middle strip. The magnetic charge density on the left and right surfaces (orange surfaces in fig. 2(c)) are $-M_s m_x$ and $M_s m_x$, respectively.

The total field from the left surface (fig.(2c)) can then be estimated by

$$H_{\text{left}} = -\frac{M_s m_x}{4\pi} \int_{-\infty}^{\infty} \int_{-\frac{d}{2}}^{\frac{d}{2}} \frac{1}{(x^2+y^2+z^2)} \frac{-x}{\sqrt{x^2+y^2+z^2}} dz dy = \frac{\arctan\frac{d}{2x}}{\pi} M_s m_x,$$

Similarly, for the right surface

$$H_{\text{right}} = -\frac{\arctan\frac{d}{2x}}{\pi} M_s m_x,$$

where $x$ is the distance from the center point to the surface.

For the $n$th strip, the distance is $x = \pm\frac{w}{2} + n\lambda$. Thus, the total dipolar field on the center of a strip reads

$$H^x_{\text{int}} = \frac{M_s m_x}{\pi} \sum_{n=-\infty}^{+\infty} \left[\arctan\frac{d}{(-w+2n\lambda)} - \arctan\frac{d}{(w+2n\lambda)}\right],$$

Since we have already considered the shape anisotropy of the center strip, the $n = 0$ term should be omitted. This term can be absorbed in $N_x$ ( $N'_x = N_x - \frac{2}{\pi}\sum_{n=1}^{+\infty}\left[\arctan\frac{d}{(-w+2n\lambda)} - \arctan\frac{d}{(w+2n\lambda)}\right]$). On the other hand, the magnetic charge density on top and bottom surfaces (blue surfaces) are $M_s m_z$ and $-M_s m_z$, respectively. The $x$ component cancels out due to the MCs symmetry, and the $z$ component at the center is

$$H^z_{\text{top}} = 2\frac{M_s m_z}{4\pi}\int_{-\infty}^{\infty}\int_{n\lambda-\frac{w}{2}}^{n\lambda+\frac{w}{2}} \frac{1}{(x^2+y^2+d^2/4)}\frac{-d/2}{\sqrt{x^2+y^2+d^2/4}}dxdy,$$

and the total $z$-componet is

$$H^z_{\text{int}} = -\frac{M_s m_z}{\pi}\sum_{n=-\infty}^{+\infty}\left[\arctan\frac{(w+2n\lambda)}{d} - \arctan\frac{(-w+2n\lambda)}{d}\right]$$

This term can be absorbed in $N_z$ ($N'_z = N_z + \frac{2}{\pi}\sum_{n=1}^{+\infty}\left[\arctan\frac{(w+2n\lambda)}{d} - \arctan\frac{(-w+2n\lambda)}{d}\right]$).

For all the MCs that we have prepared, $N_y < N'_x < N'_z$. Thus, the equilibrium magnetization tends to be in-plane and along the $y$-direction. When applying the external field along the $x$-direction, the equilibrium magnetization is tilted with respect to the $y$-direction satisfying $\boldsymbol{m} \parallel \boldsymbol{H}_{\text{eff}}$, so that

$$\frac{H - N'_x M_s \cos\phi_m}{-N_y M_s \sin\phi_m} = \frac{\cos\phi_m}{\sin\phi_m},$$

where $\phi_m$ is the azimuthal angle of $\boldsymbol{m}$. The largest $H$ for the above equation to have a solution $\phi_m = \arccos\frac{H}{(N'_x - N_y)M_s}$ is $H = (N'_x - N_y)M_s$. Above this value, $\boldsymbol{m}$ is saturated along the $x$- direction. To obtain the eigenfrequency of the MCs, we expand $\boldsymbol{m}$ around its equilibrium direction, assume the small precessional component to have a harmonic form $e^{i\omega t}$, and keep only linear terms. The result is the well-known Smit-Beljers formula[26],

$$\omega = \begin{cases} \gamma\sqrt{(N_z'M_s - N_x'M_s)\left(N_yM_s - N_x'M_s + \frac{H^2}{N_yM_s - N_x'M_s}\right)} & H < (N_x' - N_y)M_s \\ \gamma\sqrt{(N_z'M_s - N_x'M_s + H)(N_yM_s - N_x'M_s + H)} & H \geq (N_x' - N_y)M_s \end{cases} \quad (1)$$

where $\omega = 2\pi f$ is the eigenfrequency of the FMR mode. We plot the eigenfrequency versus $H$ in fig. 2(d) to understand the behavior observed in the fig. 2(a). The horizontal line at 9.4 GHz corresponds to the frequency of the cavity, and its intersections with the dispersion curve correspond to the two experimentally observed resonant modes. The low field mode is an unsaturated mode around an intermediate equilibrium magnetization state between the long and short axes of the strips. The high field mode is attributed to the uniform precession in the fully saturated state along the $x$-direction.

Concentrating on the saturated mode, the fig. 2(b) depicts the resonance field of the mode vs lattice constant. The red solid line is the result from Eq. (1) with $w = 50$ nm and fitting parameter $d = 14.0$ nm. The result agrees with the experimental data within $\pm 3\%$ error. To be more accurate, and to further verify the validity of the macrospin model, we perform micromagnetic simulations using the open-source package Mumax3[27]. The simulations were performed using $1 \times 1 \times d$ nm³ meshes. Limited by the computational capacity, the system size was set to be $1024 \times 1024 \times 1$ meshes with periodic boundary conditions. The exchange constant $A = 1.3 \times 10^{-11}$ J/m. For same geometry, the simulation results of the resonance fields are larger than that obtained by the macrospin model. This fact has also been observed in other studies[13]. Thus, to fit the experimental data, we have to use a smaller thickness $d = 12.5$ nm. The simulation results are shown by black squares in the fig. 2(b). The reason why results of analytical model based on macrospin approximation differ from the simulations is as follows. The magnetization throughout each strip is not homogeneous as assumed in the macrospin model. At the two edges, the precessional amplitude is larger, while near the center the amplitude is smaller [see inset of the fig. 2(b) as well as fig. S3(b) of the Supplemental Material]. The first consequence is that the total effective field is also inhomogeneous in the strips[28]. There is a dipolar interaction field near the center, so a larger external field is necessary to reach the resonance than in the macrospin model. The second consequence is that the demagnetization factor in the $z$-direction ($N_z'$) is significantly overestimated in the macrospin model, while $N_x'$ and $N_y$ are not affected much. Nevertheless, the macrospin model qualitatively reproduces the experimental results, showing that the main reason of the increasing resonant field is the decreasing inter-strip dipolar interaction when the separation becomes larger. To compensate the overestimated $N_z$, we introduce an empirical dimensionless parameter $\eta < 1$ to renormalize $N_z$ in Eq. (1) (for the saturated peak $H \geq (N_x' - N_y)M_s$),

$$\omega = \gamma\sqrt{(\eta N_z'M_s - N_x'M_s + H)(N_y M_s - N_x'M_s + H)}, \qquad (2)$$

For $w = 50$ nm and $d = 12.5$ nm obtained from the simulation, we find that $\eta = 0.82$ fits the simulation results best, as shown by the green dashed line in the fig. 2(b). The gray area between the two black dashed lines means the range of $H$ from Eq. (2) with $w = 50 \pm 2$ nm and $d = 12.5 \pm 0.3$ nm, in which the errors are estimated from the limited precision of lithography and the e-beam evaporation techniques. The experimental data are well in the range indicated by the gray area in the fig. 2(b).

We also prepared MCs samples with larger strip width, in which the macrospin approximation obviously fails. Fig. 3(a) shows the derivative FMR absorption of a $w = 200$ nm, $\lambda = 250$ nm sample when rotating the applied field in-plane. We can observe that for any $\phi$ angle, there is at least one resonant mode. When the field is along the $x$-direction, there are three resonant modes, instead of two in the previous samples. To understand the three modes, we performed micromagnetic simulations for $\boldsymbol{H} \parallel \hat{\boldsymbol{x}}$. After Fourier transform, the result is shown in Fig. 3(b). Three peaks can be observed, which is consistent with the experiment. More details can be found in Supplemental Material, since the main topic of the paper concerns the narrow-strip samples.

### 3.2. Magnetic Damping

To investigate the impact of dipolar coupling on magnetic damping, the recorded FMR signal was fitted to the sum of derivatives of the symmetric and antisymmetric Lorentzian functions to determine the lineshape parameters: resonance field ($H_r$) and the full width of half maxima ($\Delta H$)[22]. The fitting determined linewidth versus frequency data is fitted to the model:

$$\mu_0 \Delta H(f) = \mu_0 \Delta H_0 + \frac{4\pi \alpha_{eff}}{\gamma} f \tag{3}$$

where $\alpha_{eff}$ is the effective damping parameter including the Gilbert damping and the eddy current contributions[29] and $\Delta H_0$ is the linewidth at zero frequency also known as inhomogeneous linewidth broadening[30].

The linewidth broadening was measured for the magnetic field applied parallel and perpendicular to the strips. Table 1 summarizes the characteristics of the FMR linewidth at different frequencies when the field is applied along and perpendicular to the strips ($y$-direction) (specific data can be found in Fig. S6 in supplemental materials). In total, the effective damping parameter $\alpha_{eff} = 0.0045 \pm 0.0005$, and is almost independent of the strip separation. The inhomogeneous linewidth broadening decreases with increasing lattice constant (or strip separation). The same trend is obtained when the field is perpendicular to the strips ($x$-direction). The inhomogeneous broadening is small for a reference thin film sample without the MCs structure, which is within expectation because the thin film is much more homogeneous. An interesting observation is that the effective Gilbert damping is very different for $\mathbf{H} \parallel \hat{y}$, $\mathbf{H} \parallel \hat{x}$ and thin film experiments. It is noted that different multi-magnon scattering processes[31] and metallic electromagnetic effects such as eddy currents[32] can contribute to such effects.

### 4. Discussions and Summary

A lot of effort has been made to understand the dynamics of MCs[8,12-14,31,33,34]. For the FMR mode, one important aspect is to understand the deviation from the macrospin model. Since the dimensions of the magnetic strips are usually much larger than the exchange length, $l_{ex} = \sqrt{\frac{2A}{\mu_0 M_s^2}}$, a deviation is natural due to an inhomogeneous magnetization profile. As expected, the micromagnetic simulations shows that, when the mesh size is comparable to strip size, the macrospin model is recovered. When reducing the mesh size, the resonant field becomes lower, until the mesh size is as small as $l_{ex}$ and the result converges to the "genuine" value. This observation is also applicable when **m** is allowed to be inhomogeneous in the thickness direction. For thin strips whose thickness is much smaller than the width, an analytical formula for boundary conditions has been derived which can be used to solve the nonuniform magnetization profile[35]. However, in our case the thickness and the width are of the same order of magnitude. So, we introduce an empirical parameter $\eta$ which depends only on the bar dimensions, but independent of the bar separation, as a correction to the macrospin model. Recently, there is a very comprehensive study on the FMR mode of MCs[34]. Frequency-sweeping FMR was modeled and studied in that paper. Here, we consider field sweeping FMR, and we provide an intuitive picture for the deviation and failure of macrospin model. For 50 nm strips, the strip edge and center have different oscillation amplitudes but same phase, so the macrospin model is still qualitatively correct (i.e. emergence of two modes at two equilibrium **m** directions), and the quantitative results can be recovered by introducing a compensation factor. For 200-nm strips, the oscillations at strip edge and center are out-of-phase, so the macrospin model fails and full micromagnetic simulation is necessary.

We observe a strong anisotropy in the line broadening. Both the interception (the inhomogeneous broadening) and the slope (the effective damping) are anisotropic. The effective damping includes the Gilbert damping and dissipation by the eddy current. The Gilbert damping is usually isotropic[30], but the eddy current is anisotropic because it is related to the geometry of the sample. This is confirmed by our numerical simulation, where we assume an isotropic Gilbert damping and we do not observe the anisotropy in the effective damping. Therefore, we attribute the anisotropic effective damping to the eddy current effect[29,32]. The inhomogeneous line broadening contains the contribution from external sources such as the multi-magnon scattering, anisotropy, and scattering due to roughness and defects. These effects can be strongly anisotropic[31,36], and are not considered in the simulation. So, we suppose that they could be the reason for the observed anisotropic $\Delta H_0$. We also observe that $\Delta H$ is very small for a film but shows a decreasing trend when increasing the strip separation. This may relate to the inhomogeneity of the whole sample. The inhomogeneous broadening is positively related to the inhomogeneity of the sample[37].

In summary, we have investigated the magnetodynamic properties of 1D MCs with different lattice constants. The resonant field found to increase with increasing lattice constant because of decreasing inter-strip dipolar coupling. The experimental results are qualitatively explained by a macrospin model when the strips are narrow. The accuracy of the macrospin model can be quantitatively improved by renormalizing the out-of-plane demagnetization factor fitted by micromagnetic simulations. Obvious difference in linewidth slopes was found for different field directions.

**Acknowledgements**

This work was partly supported by the Research Council of Norway through its Centre of Excellence funding scheme, project number 262633, "QuSpin". S. S. acknowledges partial funding obtained from the Norwegian PhD Network on Nanotechnology for Microsystems, which is sponsored by the Research Council of Norway, Division for Science, under contract no. 221860/F40. The Research Council of Norway is acknowledged for the support to the Norwegian Micro- and Nano-Fabrication Facility, NorFab, project number 245963/F50. Gopal Dutt is acknowledged for the AFM measurements. X. S. W. acknowledges the support from the Natural Science Foundation of China (Grant No. 11804045).

# References

1   V. V. Kruglyak, S. O. Demokritov, and D. Grundler,  J Phys D Appl Phys **43** (26) (2010).
2   S. Neusser and D. Grundler,  Adv Mater **21** (28), 2927 (2009).
3   A. V. Chumak, A. A. Serga, and B. Hillebrands,  Nat Commun **5** (2014).
4   A. V. Chumak, A. A. Serga, and B. Hillebrands,  J Phys D Appl Phys **50** (24) (2017).
5   A. A. Nikitin, A. B. Ustinov, A. A. Semenov, A. V. Chumak, A. A. Serga, V. I. Vasyuchka, E. Lahderanta, B. A. Kalinikos, and B. Hillebrands,  Applied Physics Letters **106** (10) (2015).
6   M. P. Kostylev, A. A. Stashkevich, and N. A. Sergeeva,  Phys Rev B **69** (6) (2004).
7   G. Gubbiotti, S. Tacchi, G. Carlotti, P. Vavassori, N. Singh, S. Goolaup, A. O. Adeyeye, A. Stashkevich, and M. Kostylev,  Phys Rev B **72** (22) (2005).
8   Justin M. Shaw, T. J. Silva, Michael L. Schneider, and Robert D. McMichael,  Phys Rev B **79** (18), 184404 (2009).
9   Z. K. Wang, V. L. Zhang, H. S. Lim, S. C. Ng, M. H. Kuok, S. Jain, and A. O. Adeyeye,  Acs Nano **4** (2), 643 (2010).
10  M. Belmeguenai, M. S. Gabor, F. Zighem, D. Berling, Y. Roussigne, T. Petrisor, S. M. Cherif, C. Tiusan, O. Brinza, and P. Moch,  J Magn Magn Mater **399**, 199 (2016).
11  M. Krawczyk, S. Mamica, M. Mruczkiewicz, J. W. Klos, S. Tacchi, M. Madami, G. Gubbiotti, G. Duerr, and D. Grundler,  J Phys D Appl Phys **46** (49) (2013).
12  R. A. Gallardo, T. Schneider, A. Roldán-Molina, M. Langer, J. Fassbender, K. Lenz, J. Lindner, and P. Landeros,  Phys Rev B **97** (14), 144405 (2018).
13  M. Langer, R. A. Gallardo, T. Schneider, S. Stienen, A. Roldán-Molina, Y. Yuan, K. Lenz, J. Lindner, P. Landeros, and J. Fassbender,  Phys Rev B **99** (2), 024426 (2019).
14  S. Mamica, M. Krawczyk, and D. Grundler,  Physical Review Applied **11** (5), 054011 (2019).
15  M. Kostylev, P. Schrader, R. L. Stamps, G. Gubbiotti, G. Carlotti, A. O. Adeyeye, S. Goolaup, and N. Singh,  Applied Physics Letters **92** (13) (2008).
16  M. Krawczyk and D. Grundler,  J Phys-Condens Mat **26** (12) (2014).
17  G. Gubbiotti, S. Tacchi, M. Madami, G. Carlotti, A. O. Adeyeye, and M. Kostylev,  J Phys D Appl Phys **43** (26) (2010).
18  J. Topp, D. Heitmann, M. P. Kostylev, and D. Grundler,  Phys Rev Lett **104** (20) (2010).
19  B. Pigeau, C. Hahn, G. de Loubens, V. V. Naletov, O. Klein, K. Mitsuzuka, D. Lacour, M. Hehn, S. Andrieu, and F. Montaigne,  Phys Rev Lett **109** (24) (2012).
20  S. D. Sloetjes, E. Digernes, C. Klewe, P. Shafer, Q. Li, M. Yang, Z. Q. Qiu, A. T. N'Diaye, E. Arenholz, E. Folven, and J. K. Grepstad,  Phys Rev B **99** (6) (2019).
21  G. N. Kakazei, Y. G. Pogorelov, M. D. Costa, T. Mewes, P. E. Wigen, P. C. Hammel, V. O. Golub, T. Okuno, and V. Novosad,  Phys Rev B **74** (6) (2006).
22  M. Harder, Z. X. Cao, Y. S. Gui, X. L. Fan, and C. M. Hu,  Phys Rev B **84** (5) (2011).
23  S. Akansel, A. Kumar, N. Behera, S. Husain, R. Brucas, S. Chaudhary, and P. Svedlindh,  Phys Rev B **97** (13) (2018).
24  H. T. Nembach J. M. Shaw, T. J. Silva, and C. T. Boone,  Journal of Applied Physics **114** (2013).
25  A. Aharoni,  J Appl Phys **83** (6), 3432 (1998).
26  and H. G. Belgers J. Smit,  Philips Res. Rep. **10**, 113 (1955).
27  A. Vansteenkiste, J. Leliaert, M. Dvornik, M. Helsen, F. Garcia-Sanchez, and B. Van Waeyenberge,  Aip Adv **4** (10) (2014).
28  C. Bayer, J. Jorzick, S. O. Demokritov, A. N. Slavin, K. Y. Guslienko, D. V. Berkov, N. L. Gorn, M. P. Kostylev, and B. Hillebrands,  Spin Dynamics in Confined Magnetic Structures Iii **101**, 57 (2006).

[29] D. S. Chrischon, F. Beck, K. D. Sossmeier, and M. Carara, Journal of Magnetism and Magnetic Materials **336**, 66 (2013).
[30] J. M. Shaw, T. J. Silva, M. L. Schneider, and R. D. McMichael, Phys Rev B **79** (18) (2009).
[31] I. Barsukov, F. M. Romer, R. Meckenstock, K. Lenz, J. Lindner, S. H. T. Krax, A. Banholzer, M. Korner, J. Grebing, J. Fassbender, and M. Farle, Phys Rev B **84** (14) (2011).
[32] V. Flovik, F. Macia, A. D. Kent, and E. Wahlstrom, J Appl Phys **117** (14) (2015).
[33] Rodrigo Arias and D. L. Mills, Phys Rev B **60** (10), 7395 (1999); J. Ding, M. Kostylev, and A. O. Adeyeye, Phys Rev B **84** (5), 054425 (2011).
[34] G. Centała, M. L. Sokolovskyy, C. S. Davies, M. Mruczkiewicz, S. Mamica, J. Rychły, J. W. Kłos, V. V. Kruglyak, and M. Krawczyk, Phys Rev B **100** (22), 224428 (2019).
[35] K. Yu Guslienko and A. N. Slavin, Phys Rev B **72** (1), 014463 (2005).
[36] W. K. Peria, T. A. Peterson, A. P. McFadden, T. Qu, C. Liu, C. J. Palmstrøm, and P. A. Crowell, Phys Rev B **101** (13), 134430 (2020).
[37] Ernst Schlömann, Physical Review **182** (2), 632 (1969).

**Table 1** The Gilbert damping parameters estimated for MCs and the reference Py thin film

| Sample | H along y-axis | | H along x-axis | |
|---|---|---|---|---|
| | α | ΔH (Oe) | α | ΔH(Oe) |
| $\lambda = 100$ nm, $w = 50$ nm | 0.004 | 97 | 0.010 | 72 |
| $\lambda = 150$ nm, $w = 50$ nm | 0.004 | 58 | 0.012 | 35 |
| $\lambda = 250$ nm, $w = 50$ nm | 0.005 | 37 | - | - |
| Py thin-film | 0.007 | 5 | 0.007 | 5 |

**Figure captions**

**Fig. 1** (a) SEM image of a MC with $\lambda = 100$ nm and $w = s = 50$ nm. (b) Schematic diagram of the coordinate system defining geometry of the field.

**Fig. 2** (a) The polar plot of resonant field extracted for samples of different $\lambda$. (b) The measured higher resonant fields when the applied field is along x direction for different samples (blue dots). The red line is the macrospin model result (Eq. 1) for $d = 14.0$ nm and $w = 50$ nm. The simulation results for $d = 12.5$ nm and $w = 50$ nm are shown as black squares. The green dashed line is effective model (Eq. 2) for $d = 12.5$ nm and $w = 50$ nm with empirical coefficient $\eta = 0.82$. The gray area indicates the range of resonant field for $d = 12.5 \pm 0.3$ nm and $w = 50 \pm 2$ nm. The inset schematically illustrates the amplitude of magnetization tilting during the precession for the macrospin model and the actual simulation. (c) Schematics depicting the geometry of magnetic strips used in analytical calculations and micromagnetic simulations. (d) The dispersion curves for $\lambda = 100$ nm and $\lambda = 550$ nm. The dashed horizontal line is the cavity frequency 9.4 GHz

**Fig. 3** (a) Experimental results of angle ($\phi$)-dependent differential FMR absorption for the $\lambda = 250$ nm, $w = 200$ nm sample at RT. (b) Fourier amplitude of the average magnetization estimated by micromagnetic simulations on the $\lambda = 250$ nm, $w = 200$ nm sample for field along $\phi = 0$. Three peaks can be identified, which is consistent with the experiment (indicated by the arrows).

(a) 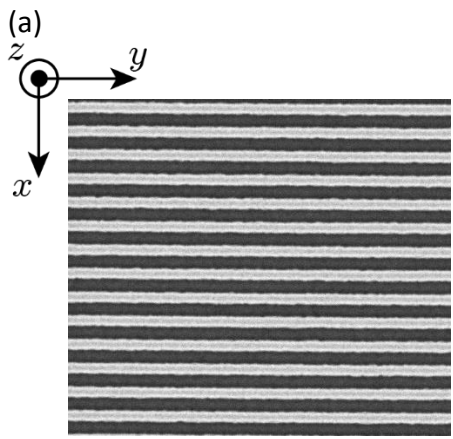

(b) 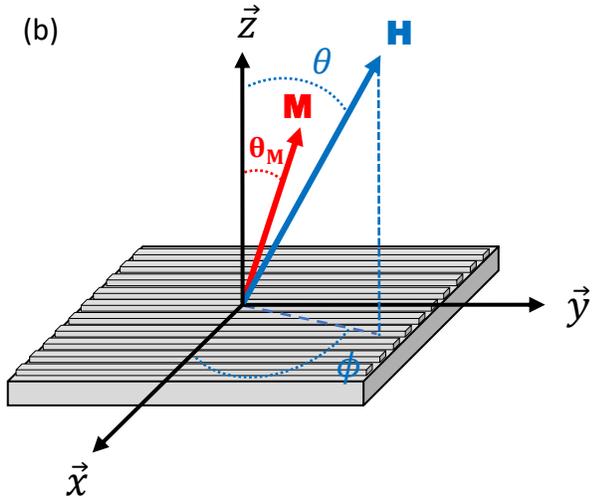

**Fig. 1**

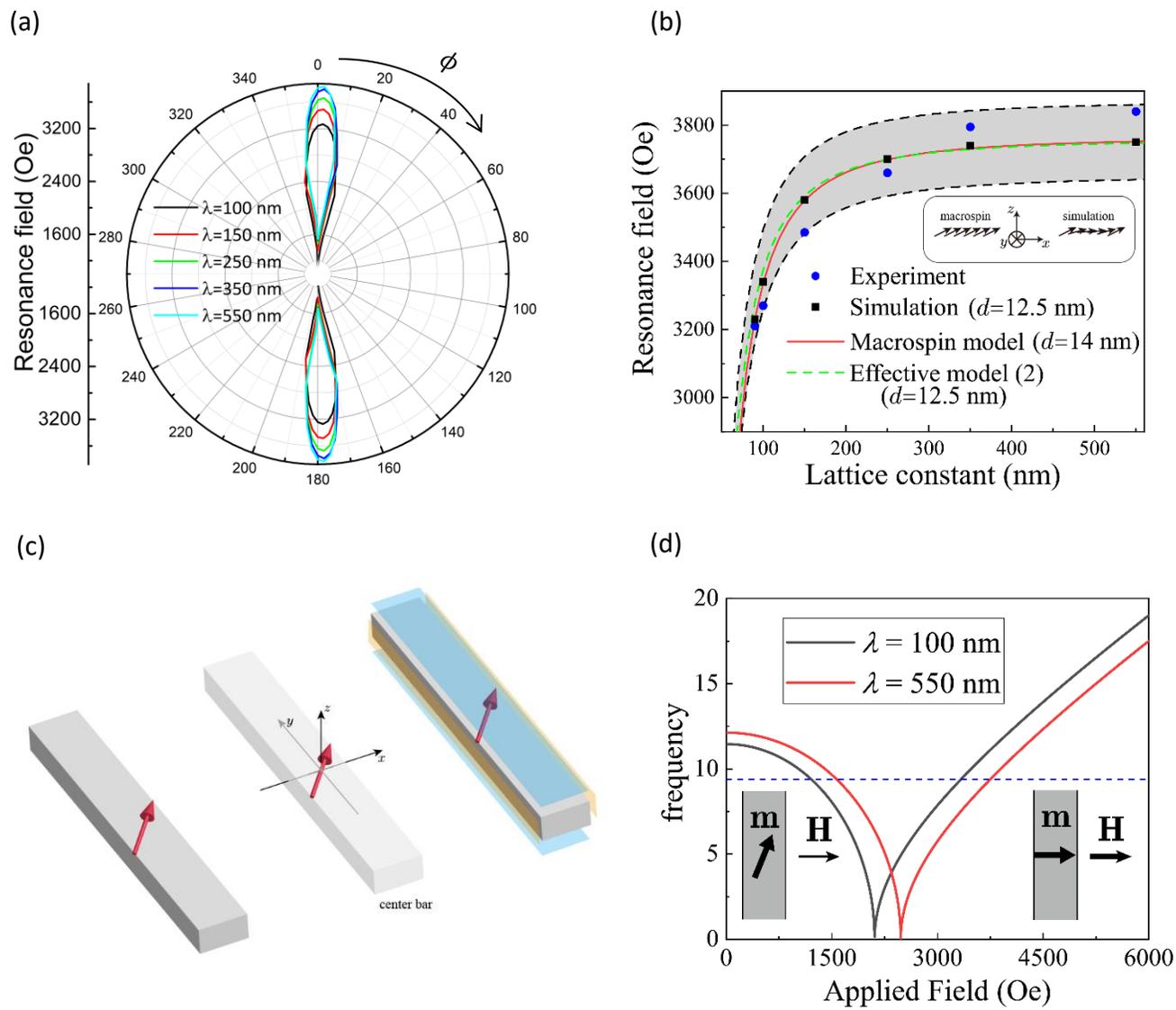

**Fig. 2**

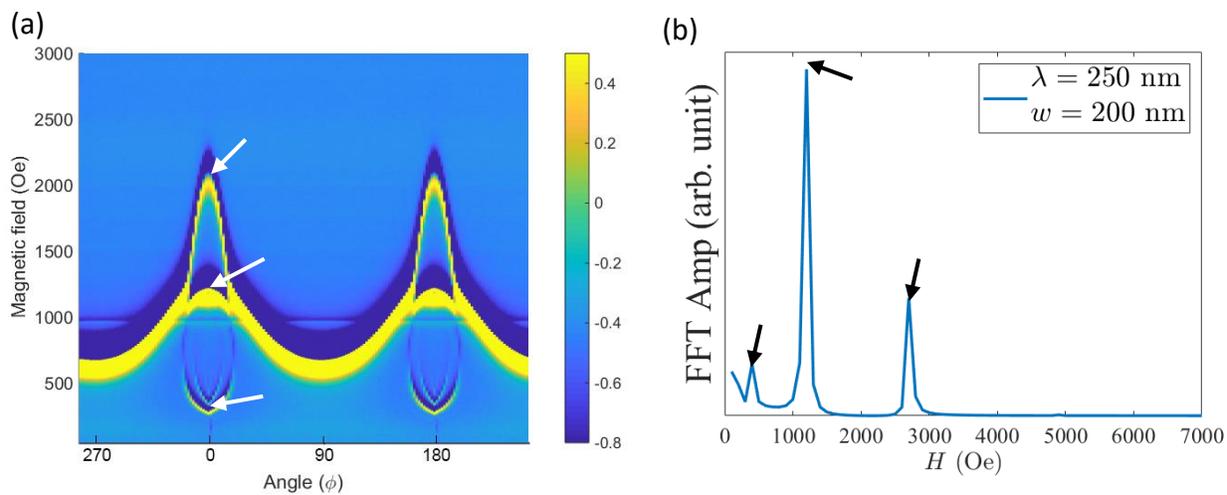

**Fig. 3**

# Supplementary Information

# Mapping the dipolar coupling and magnetodynamic properties of dipole coupled 1D magnonic crystals

Suraj Singh[1], Xiansi Wang[1], Ankit Kumar[2], Alireza Qaiumzadeh[1], Peter Svedlindh[2], Thomas Tybell,[3] and Erik Wahlström[1]

[1]Center for Quantum Spintronics, Department of Physics, NTNU - Norwegian University of Science and Technology, NO-7491 Trondheim, Norway

[2]Department of Materials Sciences and Engineering, Uppsala University, Box 516, SE-75121 Uppsala, Sweden

[3]Department of Electronic Systems, NTNU - Norwegian University of Science and Technology, NO-7491 Trondheim, Norway

Figure S1(a) shows the raw cavity FMR spectra recorded on the magnonic crystals. Two modes were observed for the magnetic applied perpendicular to strips. Fig. S1(b) shows the angle-dependent FMR measurements. The two modes found shifting towards the each other as the magnetic field is rotated away from the angle (ɸ = 0°). The resonance field of low field mode increases as a function of angle (ɸ) whereas the it decreases for the high field mode. The modes merge into a single-mode at around ɸ= ±15° and then disappears.

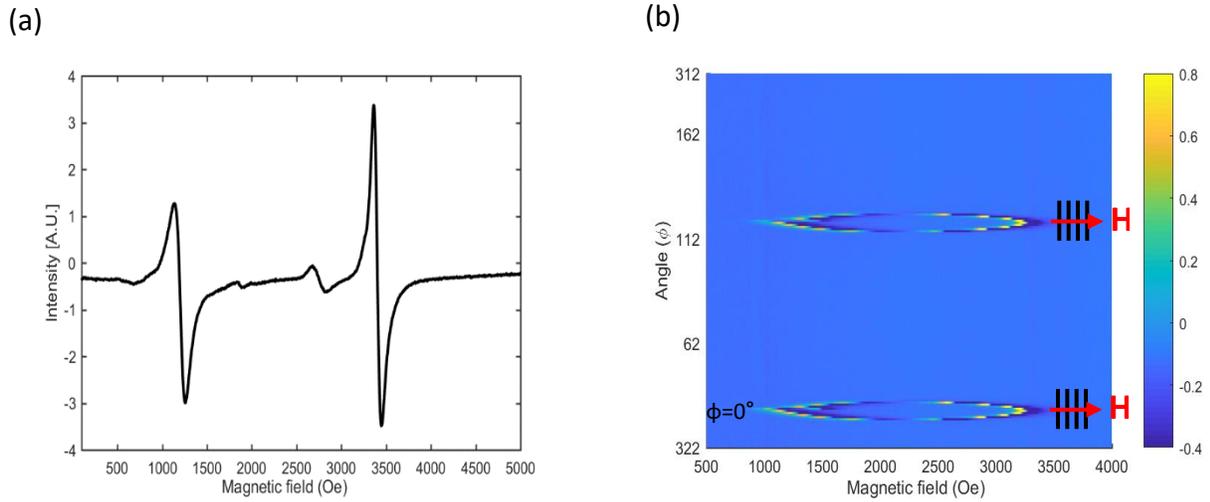

**Fig. S1** (a) The FMR lineshape recorded on $\lambda = 100$ nm and $w = 50$ nm sample for the magnetic field applied perpendicular to the strips. S2(b) The angle-dependent FMR spectra recorded rotating the magnetic field in the plane of the sample from angle ɸ = 0° to 360°.

The micromagnetic simulations performed on the magnonic lattices were used to investigate the equilibrium magnetization profile at the observed modes. Figures. S2(a) and S2(b) depict the equilibrium magnetization profiles of the low-field mode and the high-field mode observed in $\lambda = 100$ nm, $w = 50$ nm sample for the magnetic field applied along x-axis. As described in the main paper, the high field mode occurs at $H = 3.34 \times 10^3$ Oe when the magnetization is homogeneously magnetized along the x direction. The low field occurs at $H = 1 \times 10^3$ Oe when $\boldsymbol{m}$ is oriented about 63° with respect to the x-axis in the bars . Although $\boldsymbol{m}$ is not uniform over the bar, the variation is not large.

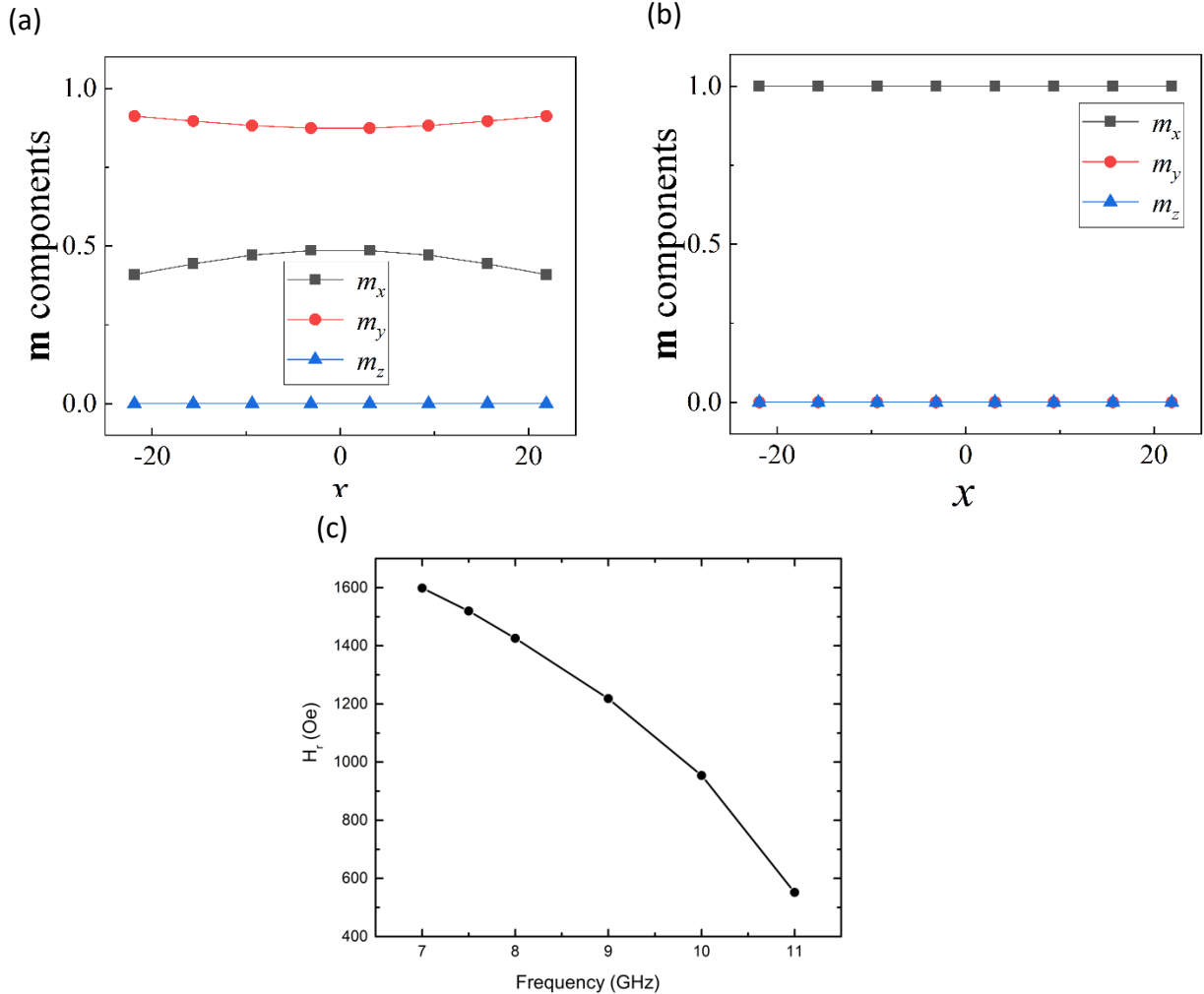

**Fig. S2** The equilibrium profile of static magnetization at the low field mode in (a) and high field mode in (b) ) for $\lambda = 100$ nm, $w = 50$ nm sample. (c) The resonant field plotted as function frequency for the low field mode.

Knowing the equilibrium magnetization profile, time-dependence of the $m_z$ component at resonance and its behaviors at the edge and the center of the strip can be plotted, see figs. S3(a) and S3(b). It can be seen that the edge and the center precess in-phase at both peaks. At the first peak, the amplitudes at the edge and center are similar, while at the second peak the edge has a larger amplitude than the center.

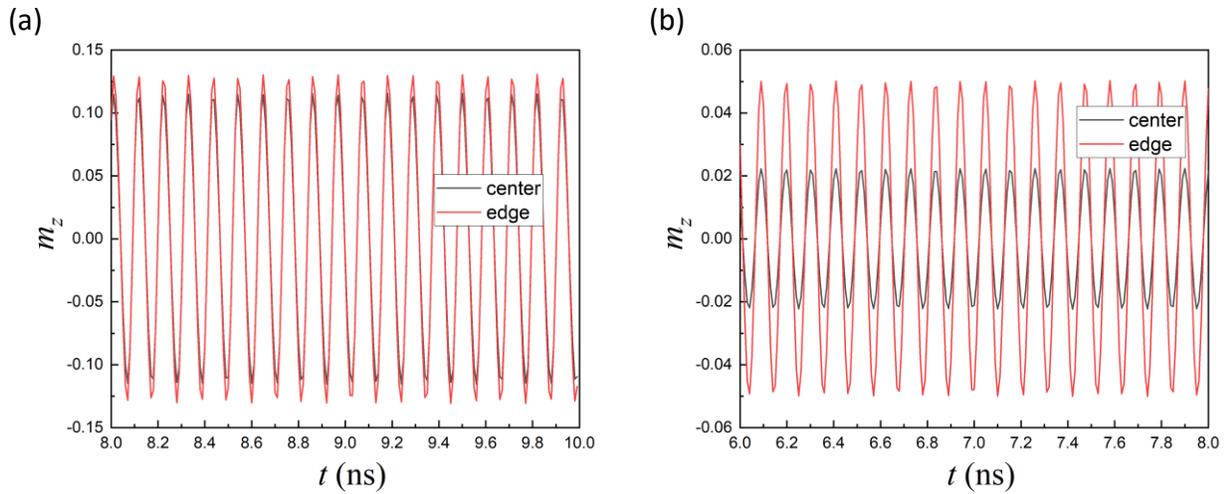

**Fig. S3** The amplitude of the magnetization precession at the low field mode in (a) and at high field mode in (b).

The equilibrium magnetization profiles at the three peaks from low field to high field for the $\lambda = 250$ nm, $w = 200$ nm sample for the magnetic field applied along x-axis are shown in figs. S4(a), S4(b), and S4(c). The magnetic fields are at the resonant peaks shown in Fig. 3(b) in the main text. The magnetization pofile is much different than the $\lambda = 100$ nm, $w = 50$ nm sample because the bars are much wider and the spatial variation of $\boldsymbol{m}$ is much more significant in the two low-field peaks. The time dependence of $m_z$ oscillation from the three peaks is shown figs. S5(a), S5(b), and S5(c).

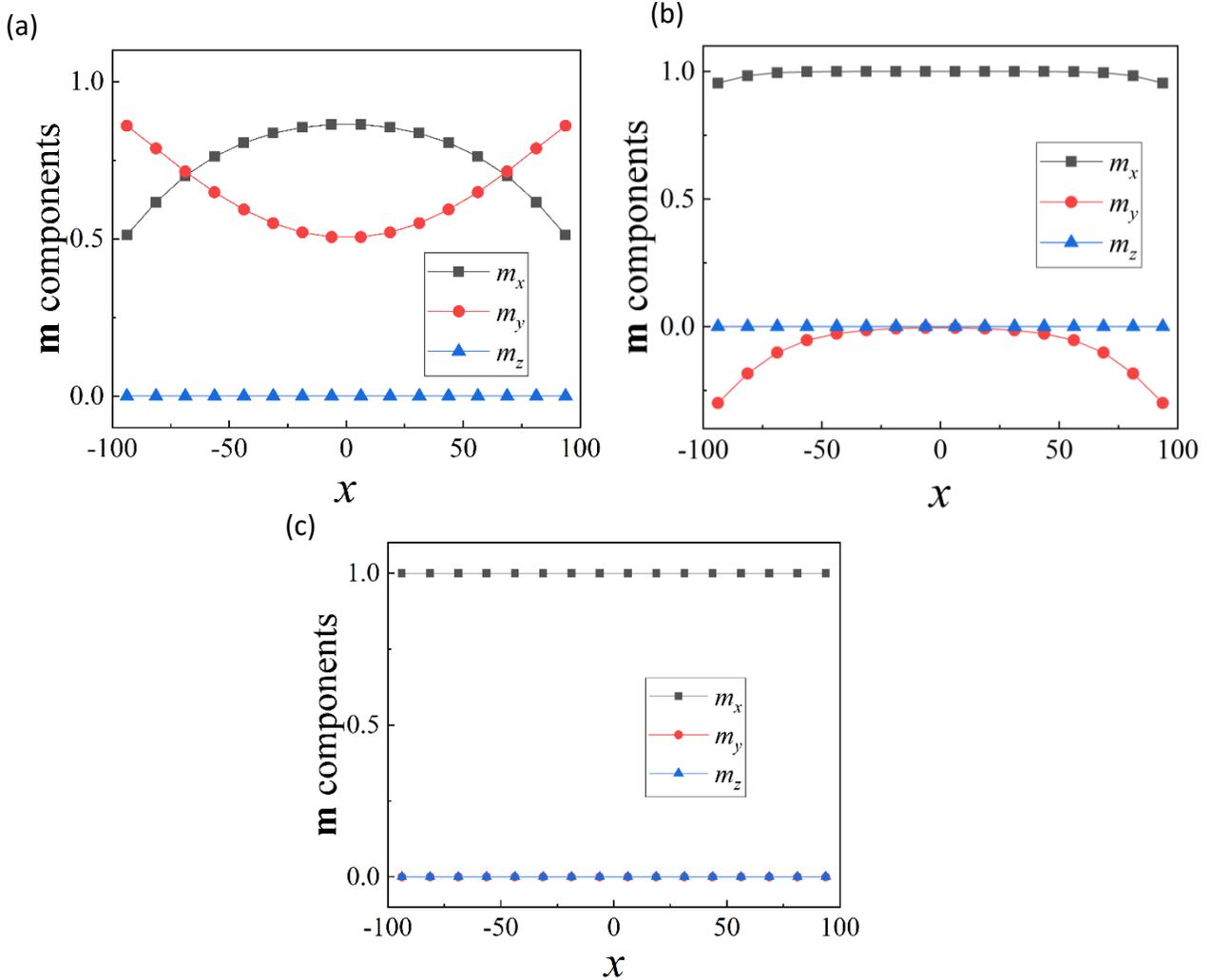

**Fig. S4** The equilibrium magnetization profiles at three peaks observed in $\lambda = 250$ nm, $w = 50$ nm sample from low to high in (a), (b), and (c) respectively.

At the first peak, the edge and the center have a $\pi$-phase difference. Since the energy absorption is measured, the energy absorptions at the edge and the center cancel with each other. So although the oscillation amplitude is large, the peak intensity is small. At the second peak, the center magnetization is almost along the x direction, but edge magnetization is significantly tilted. The oscillations also have a $\pi$-phase difference, but at the center the amplitude is much larger. Furthermore, the center magnetization has a larger proportion in the whole bar. So the energy absorption is dominated by the central part, and has the largest intensity. At the third peak, the magnetization is fully aligned along the x direction. The oscillations have a $\frac{\pi}{2}$-phase difference, but the edge has a much larger amplitude. The energy absorption is dominated by the edge part.

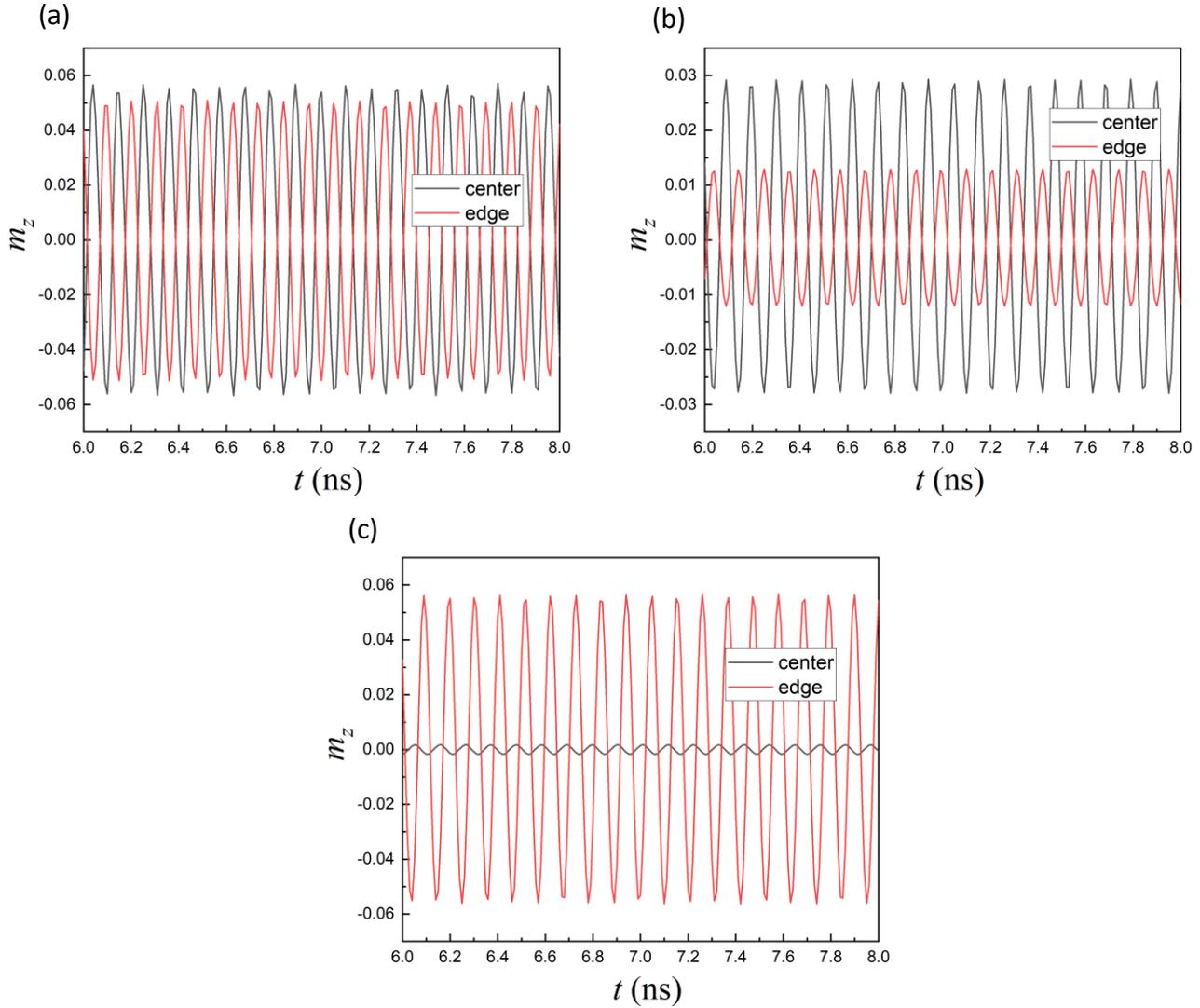

**Fig. S5** The amplitude of the magnetization precession at the three peaks from low to high in (a), (b) and (c) respectively.

For both $\lambda = 100$ nm, $w = 50$ nm and $\lambda = 250$ nm, $w = 200$ samples the amplitudes at the edges are larger than that at the center for the saturated peaks. But for other peaks the comparison is complicated. This can be understood as follows. At equilibrium, the local effective field is parallel to the local magnetization for each point. The larger magnitude the effective field has, the harder the magnetization precesses. For the saturated peaks, the edges bear larger demagnetization field (because they are closer to the surface magnetic charges). The demagnetization field is antiparallel to $m$, so the total effective field at the edges is smaller than the center. For other non-saturated peaks, the situation is very complicated, so there is no simple picture for the observed amplitude difference.

The details of the measurements of magnetic damping are summarized in fig. S6.

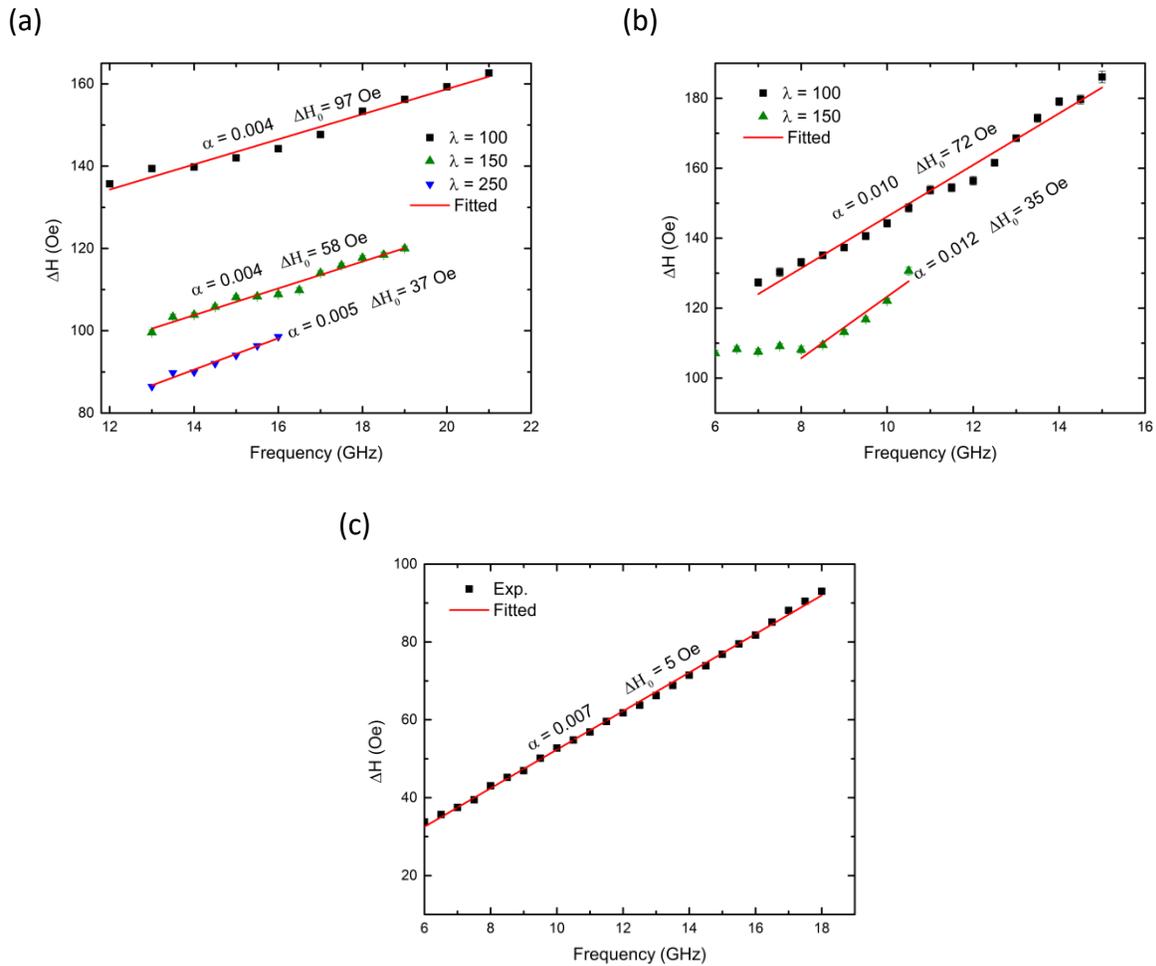

**Fig.** S6 The linewidth versus frequency data fitted to a straight line for the magnetic field applied along and perpendicular to the magnetic strips for $\lambda$ with $w = 50$ nm samples in 5(a) & 5(b) and for the reference thin-film in fig. 5(c). The filled colored symbols the experimental data points whereas the solid red line shows the fitted data.